# EEG Based Decoding of the Perception and Regulation of Taboo Words

Parisa Ahmadi Ghomroudi[*,1], Michele Scaltritti[1], Bianca Monachesi[1], Peera Wongupparaj[2]

Remo Job[1], & Alessandro Grecucci[1,3]

[1]DiPSCo – Department of Psychology and Cognitive Sciences, University of Trento, Rovereto, Italy

[2]Department of Psychology, Faculty of Humanities and Social Sciences, Burapha University, Thailand

[3]CISMed – Center for Medical Sciences, University of Trento, Trento, Italy

**Corresponding author:**

Parisa Ahmadi Ghomroudi

Department of Psychology and Cognitive Sciences,

University of Trento,

Corso Bettini, 84, 38068,

Rovereto, Italy

**E-mail**: p.ahmadighomroudi@unitn.it

**Tel.** +39 0464 808302

**Abstract**

In daily interactions, emotions are frequently conveyed and triggered through verbal exchanges. Sometimes, we must modulate our emotional reactions to align with societal norms. Among the emotional words, taboo words represent a specific category that has been poorly studied. One intriguing question is whether these word categories can be predicted from EEG responses with the use of machine learning methods. To address this question, Support Vector Machine (SVM) was applied to decode the word categories from Event Related Potential (ERP) in 40 native Italian speakers. 240 neutral, negative and taboo words were used to this aim. Results indicate that the SVM classifier successfully distinguished between the three-word categories, with significant differences in neural activity ascribed to the late positive potential mainly detected in the central-parietal-occipital and anterior right scalp areas in the time windows of 450-649 ms and 650-850 ms. These findings were in line with the established distribution pattern of the late positive potential. Intriguingly, the study also revealed that word categories were still detectable in the regulate condition. This study extends previous results on the domain of the cortical responses of taboo words, and how machine learning methods can be used to predict word categories from EEG responses.

**Introduction**

In everyday interactions, we frequently communicate our feelings through language, with emotional responses commonly triggered by verbal interactions. This dynamic is mirrored in psychotherapy sessions, where words are fundamental. Patients use words to convey their emotional experiences, and their therapists use words to mirror, elaborate, and regulate their clients' emotions[1]. An interesting class of words that convey emotions is that of taboo words. They offer unique psychological and social features, unlike any other word category. Swearing triggers emotional responses[2], enhances the impact of expressions[3], and generates humor[4]. It also regulates emotions and alleviates pain[5] and boosts the effectiveness of our messages[6],

Despite the existing knowledge on emotional words, the investigation of taboo words remains underrepresented in EEG research[7]. Existing evidence indicates that the late positive potential (LPP) is the most frequently observed ERP component associated with emotional words. This component appears as a positive deflection in the waveform around 400 ms after stimulus onset, primarily in the centro-parietal region[8,9]. Numerous studies show that emotional words elicit a stronger LPP than neutral ones, with positivity peaking at centro-parietal sites and potentially shifting toward frontal areas over time[9–16] Specifically, Kanske & Kotz[13] showed that negative words elicited a greater LPP amplitude than either neutral or positive words. However, we still do not know whether taboo words elicit comparable or different activity compared to neutral and negative words. Thus, the first aim of this study is to explore the psychological and neural differences of taboo words compared to negative and neutral words. To detect these differences, we aim to use multi-voxel pattern analysis, also known as machine learning methods, to test the hypothesis that LPP can be used to predict word categories.

A related issue concerns how we process and respond to particular categories of words, in particular with respect to their affective connotation. In daily life, we constantly modulate our emotional reactions to comply with societal norms. Emotion regulation is crucial for health and well-being [17] and it plays an important role in various psychotherapeutic approaches, including cognitive behavioral therapy[18] and acceptance and commitment therapy [19,20]. Deficits in emotion regulation are associated with a spectrum of mental health issues[21,22]. Given the significance of emotion regulation techniques in psychotherapy and the role of language in triggering emotionally charged events, it is essential to explore how emotion regulation strategies influence linguistic stimuli. In terms of neurophysiological underpinnings, the LPP component seems crucially associated with the intentional access and processing of emotional content (e.g.,[9,23,24]). This characteristic makes it a perfect component for assessing the impact of emotion regulation strategies[25].

Several studies have examined how emotion regulation strategies influence the processing of negative and positive word stimuli. Deveney & Pizzagalli[26] found changes in brain activity (N400 and P300 amplitudes) when participants used reappraisal to regulate emotions after viewing unpleasant images. However, this latter study did not directly show how emotion regulation affects word processing. Baker et al[27] explored three strategies—compassion-focused reappraisal, benefit-focused reappraisal, and offense rumination—finding that benefit-focused reappraisal increased emotional response to unpleasant words, while the other strategies had no effect. Grecucci et al[1]. found that distancing increased brain activity in response to unpleasant pictures but not in response to word stimuli. Most ERP studies have focused on reappraisal's impact on emotional pictures. In the present study we aim to extend previous results to a less studied strategy known as acceptance. Acceptance, an emotion regulation strategy defined by a non-judgmental and open stance toward emotional experiences, has been shown to improve mental health and reduce emotional reactivity and physiological responses to distressing emotions. Studies suggest its effectiveness in managing emotions [12,28]. Of note, although the previous studies showed a reduced neural difference between word categories in the regulate condition, and we expect to observe a similar reduction, it remains unclear whether word categories can be still distinguished at a neural level. In other words, whether the remaining information can still be used to differentiate across word categories. Machine learning methods have been shown to outperform univariate statistical approaches in detecting more subtle patterns. Thus, the second aim of this study is to distinguish between categories (neutral vs taboo vs negative words) during the regulation condition. Although we expect reduced differences between the word categories, we predict that machine learning methods will still be able to detect them.

To address these aims we relied on EEG and used neutral, negative and taboo words as stimuli and asked participants to Look or regulate their emotional reactions to words. A machine learning method known as Support Vector Machine was used to this aim. Traditionally, as briefly reviewed above, univariate ERP methods have been used extensively to explore the effects of different emotion regulation strategies on both pictorial and word stimuli[29,30]. In contrast, decoding and multivariate pattern analysis (MVPA) present complementary and alternative methodologies to the traditional univariate ERP analysis[31].. Widely used in MRI and fMRI research[32,33], MVPA has also been recently used for EEG time-series data analysis[34] . Unlike univariate methods that consider voxels in fMRI, or channels in MEG/EEG as independent measures of activity, MVPA evaluates the interrelationships among these elements, thereby enhancing sensitivity to differences across experimental conditions[31] . Classifiers in decoding techniques are particularly adept at uncovering subtle information in data that univariate methods, which focus on averaged signals, might miss [31,35]. Furthermore, while univariate methods rely on activation data or regions of interest, MVPA is driven by patterns of activity across multiple brain regions or channels. In the present study, MVPA was applied to decode emotional categories from ERP responses.

In sum, the primary aim of our research was to decode affective word categories from ERPs. SVM and three contrasts will be used to this aim: neutral versus taboo, neutral versus negative, and taboo versus negative, during stimulus observation. Behaviorally, we hypothesize that neutral stimuli will be rated higher in valence compared to both negative and taboo stimuli, while taboo and negative stimuli will elicit higher arousal responses than neutral stimuli. We expect ERP activity distinguishing these word categories to occur primarily in the parieto-temporal regions, around 450 ms post-stimulus onset, a time frame typically associated with the LPP[1].. The second aim of the study was to investigate whether the same machine learning based ERP classification could be also achieved for the same word categories in the regulate condition. Although several studies have demonstrated a reduction in cortical activity during emotion regulation strategies[36,37], we hypothesize that it will still be possible to differentiate between the three word categories (neutral, taboo, and negative) even in the regulate condition. Last but not least, we aim to examine the effect of the strategy e.g., the neural activity associated with the implementation of the strategy. Previous studies, such as Grecucci et al[1]., have shown that strategies like distancing and reappraisal, are visible at the cortical level (e.g., stimulus preceding negativity). However, in another line of research by Monachesi et al [38] and Messina et al [39] have shown that acceptance does not recruit top-down cortical activity but rather bottom-up subcortical activity, as compared to more cognitive strategies. Hence, we expect no neural difference between regulate and Look, thus classifier will fail to distinguish the two conditions based on cortical activity as measured by EEG.

**Results**

**Behavioral Results**

*Word Categories in the Look Condition:*

Neutral stimuli had significantly higher valence ratings than taboo ones, $t(34) = 5.79$, $p < .001$, with a mean difference of 0.92. Similarly, valence ratings were higher for neutral compared to negative stimuli, $t(34) = 9.92$, $p < .001$, with a mean difference of 1.93. When comparing valence ratings of negative and taboo stimuli, a significant difference was observed, $t(34) = -13.19$, $p < .001$, with a mean difference of -1.01, indicating taboo stimuli were significantly higher.

In terms of arousal, a significant difference was found between neutral and taboo stimuli, $t(34) = -8.65$, $p < .001$, with a mean difference of -1.44. A significant difference was also found between neutral and negative stimuli, $t(34) = -6.32$, $p < .001$, with a mean difference of -1.61, indicating that negative stimuli were rated as more arousing than neutral stimuli. However, no significant difference was found between arousal ratings of negative and taboo stimuli, $t(34) = 1.11$, $p = .275$, with a mean difference of 0.16.

*Word Categories in the Regulation Condition*

A significant difference in valance was found between neutral and taboo stimuli, $(34) = 5.37$, $p < .001$, with a mean difference of 0.9, indicating that taboo stimuli were rated significantly lower in valence compared to neutral stimuli. A significant difference in valence was also found between neutral and negative stimuli, $t(34) = 8.31$, $p < .001$, with a mean difference of 1.88, showing neutral stimuli were significantly higher in valence ratings. When comparing valence ratings of negative and taboo stimuli, the contrast revealed a mean difference of -0.98, $t(34) = -7.80$, $p < .001$, indicating that taboo stimuli were rated significantly higher, $p < .001$.

For arousal, a significant difference was found between neutral and negative stimuli, $t(34) = -6.55$, $p < .001$, with a mean difference of -1.52, indicating negative stimuli obtained significantly higher ratings. No significant difference was found between arousal ratings of negative and taboo stimuli, $t(34) = 1.12$, $p = .27$, with a mean difference of 0.15. See further details in Table 3.

………………………………………………..

**Please insert Table 3 about here**

…………………………………

**EEG Results**

*Word Categories in the Look Condition*

In the Look condition, the model achieved a BAC of 70.0% when distinguishing between neutral and taboo stimuli, with a sensitivity of 74.3%, specificity of 65.7%, PPV of 68.4%, NPV of 71.9%, and an AUC of .80 ($p$ = .01). The confusion matrix indicated correct predictions for neutral words 74.3% of the time and taboo words 65.7% of the time. Notably, significant predictors included Region 10 in the central-parietal-occipital area and the 637-878 ms time window, as well as Region 5 in the central scalp area during the 650-850 ms time window, both showing mean feature weights greater than ±0.2.

Furthermore, the model distinguished between neutral and negative stimuli with a BAC of 54.3%, sensitivity of 57.1%, specificity of 51.4%, PPV of 54.1%, NPV of 54.5%, and AUC of .60 (p = 0.0316). The confusion matrix showed neutral words were predicted correctly 57.1% of the time, while negative words were predicted correctly 51.4% of the time. Key features included Regions 10 and 3, located in the anterior right part of the scalp during the 450-649 ms time window.

Finally, the SVM model reported a BAC of 65.7% when distinguishing negative from taboo stimuli, with a sensitivity of 74.3%, specificity of 57.1%, PPV of 63.4%, NPV of 69.0%, and AUC of .69 ($p$ = .01). According to the confusion matrix, taboo words were accurately predicted 74.3% of the time. Significant distinguishing features included Region 6 in the central right part during the 450-649 ms time window, and Region 10, with mean feature weights greater than ±0.2 (see Figures 4 and 5).

…………………………………………..

Please insert Figure 4 about here

………………………………

…………………………………………..

Please insert Figure 5 about here

………………………………

*Word Categories in the Regulation Condition*

For what concerns the Accept condition, the SVM model successfully differentiated between neutral and taboo stimuli, with a balanced accuracy of 68.6% (Sensitivity 68.6%; Specificity 68.6%; PPV 68.6%; NPV 68.6%; AUC .74; *p* = .01). The confusion matrix showed equal prediction accuracy for both neutral and taboo words at 68.6%. Key predictors of model performance included the central-right regions during the 450-649 ms window, the frontal-right region within 650-850 ms, region 10 in the central-parietal-occipital area during 637-878 ms, and the central region between 650-850 ms, all showing significant mean feature weights above ±0.2. Regarding the comparison between neutral and negative words, the model achieved a balanced accuracy of 62.9% (Sensitivity 65.7%; Specificity 60.0%; PPV 62.2%; NPV 63.6%; AUC .58; *p* = .01). The confusion matrix reveals that the classifier accurately predicted neutral and negative words 65.7% and 60% of the time, respectively. Notably, region 6 in the central-right area during the 450-649 ms window was a significant feature, with mean feature weights above ±0.2.

When differentiating between negative and taboo words, the SVM model reported a BAC of 57.1% (Sensitivity 51.4%; Specificity 62.9%; PPV 56.4%; NPV 57.7%; AUC .61; *p* = .01). The confusion matrix indicated that taboo words were predicted with 62.9% accuracy. Influential features in this analysis were region 3 between 650-850 ms and region 10, which demonstrated mean feature weights exceeding ±0.2. See further details in Table 4 Figure 6 and 7.

The SVM model differentiated between the Look and Accept conditions independently of stimuli type. However, it showed a reduced balanced accuracy of 34.3%, with a sensitivity of 24.3%, specificity of 45.7%, PPV of 36.2%, NPV of 37.2%, and AUC of .22. These performance metrics indicate suboptimal differentiation (see Figure 8).

………………………………………………..

Please insert Table 4 about here

…………………………………………….. 

……………………………………………..

Please insert Figure 6about here

……………………………………………..

……………………………………………..

Please insert Figure 7 about here

……………………………………………..

……………………………………………..

Please insert Figure 8 about here

……………………………………………..

**Discussion**

The primary aim of this study was to decode word categories from ERP activity across three contrasts: neutral versus taboo, neutral versus negative, and taboo versus negative, during stimulus observation (Look). The second aim was to investigate whether a similar ERP classification could be achieved in the regulate condition (Accept). We assessed both behavioral and neural responses to these categories. Additionally, we wanted to test the hypothesis that acceptance strategy could not be distinguished from the Look condition via cortical activity. This was motivated by the fact that previous studies on acceptance have shown little or no cortical activity[38].

As a manipulation check, the behavioral analysis confirmed significant differences in both valence and arousal between the three words categories. In the Look condition, neutral stimuli were rated significantly higher in valence than both negative and taboo stimuli, while the valence of negative stimuli was rated significantly lower than that of taboo stimuli. Arousal ratings indicated that both taboo and negative stimuli elicited a higher arousal response compared to neutral stimuli.

However, there was no significant arousal difference between negative and taboo stimuli. In the Accept condition, neutral stimuli maintained higher valence ratings compared to both negative and taboo stimuli, and the valence of negative stimuli remained significantly lower than for taboo stimuli. Similarly, arousal ratings for negative stimuli were higher than those for neutral stimuli. As in the Look condition, no significant difference in arousal was observed between negative and taboo stimuli.

When it came to the neural level, the hypothesis that word categories could be correctly classified using machine learning methods was confirmed. SVM classification (a multivariate pattern analysis decoding technique) was used to predict word categories within specific contrasts (neutral versus taboo, neutral versus negative, and taboo versus negative) from ERPs. In the Look condition, the model achieved its highest balanced accuracy of 70.0% when distinguishing between neutral and taboo words. Key contributing features were observed in the central-parietal-occipital regions during the 637-878 ms time window and in the central area between 650-850 ms. The model's ability to differentiate between neutral and negative stimuli was moderate, with a BAC of 54.3%, where the central-parietal-occipital region contributed during the 637-878 ms window and the anterior right region contributed during the 450-649 ms window. Additionally, when distinguishing between negative and taboo stimuli, the model achieved a BAC of 65.7%, with contributions from the central-right area between 450-649 ms and the central-parietal-occipital area in the 637-878 ms time window.

This indicates that, although both categories elicit strong emotional reactions, the processing of negative and taboo content differs in important ways. This may reflect the LPP's response to the negative valence of taboo words. In the Accept condition, the model successfully decoded the same word contrasts. Additionally, broader brain regions, particularly the right frontal region, were engaged. This region is commonly associated with processing negative valence and evaluating emotionally charged stimuli. For the comparison between neutral and negative words, the model reported a balanced accuracy of 57.1%, reflecting a moderated LPP response under the Accept condition. This suggests more controlled processing of negative valence, evidenced by significant activity in the central right brain region within the 450-649 ms window, indicating an adjusted neural response to negative stimuli when regulated[11].

This high accuracy likely reflects the increased LPP amplitude in response to emotionally charged words, as well as the social dimension of taboo words[23,40]. The increased LPP responses to taboo words, compared to neutral ones, were significantly influenced by neural activity in the central-parietal-occipital region during the 637-878 ms and 650-850 ms time windows post-stimulus. These findings align with the established centro-parietal distribution and peak timing of the LPP, suggesting that the neural response increases with the emotional valence of words, displaying larger amplitudes for emotionally charged stimuli compared to neutral ones[11,13,41]..

As an additional hypothesis, we predicted that the SVM could significantly classify (although with a reduced performance) the word categories in the Accept condition too. This was confirmed by our results. In the Accept condition, the model's performance in distinguishing between neutral and taboo stimuli achieved a BAC of 68.6%. The key regions contributing were the central-right area during the 450-649 ms window, the frontal-right region within the 650-850 ms window, and the central-parietal-occipital area during the 637-878 ms time window. For the neutral vs. negative contrast, the model achieved a BAC of 62.9%, with features drawn primarily from the central-right area during the 450-649 ms window. Finally, for the negative vs. taboo comparison, the model yielded a BAC of 57.1%, with features in the anterior right region during the 650-850 ms window and in the central-parietal-occipital area during the 637-878 ms time window.

Finally, consistent with our hypothesis, the SVM model performed poorly in differentiating between the Look and Accept conditions, achieving a non-significant low balanced accuracy of 34.3%. This poor performance likely reflects that the acceptance strategy relies more on bottom-up processing and subcortical activity, rather than high-level executive processes[39,42–44]. This reliance on subcortical mechanisms may explain the classifier's difficulty in differentiating between the Look and Accept conditions.

In this study, we found that taboo words, despite having higher valence scores (4.25 ± 1.03, 5.01 ± 0.58) than negative words (2.19 ± 0.37, 6.32 ± 0.57), showed increased ERP responses and higher accuracy across both conditions. This notable outcome may be attributed to the unique aspect of tabooness[45–47]. Beyond valence and arousal, tabooness introduces a dimension of social and contextual relevance, making these words more prominent and memorable during cognitive processing. The processing of taboo words may trigger specifically attentional capture phenomena and engage cognitive resources, not just due to their emotional impact but also because of their social inappropriateness or perceived threat. This supports the idea that taboo words are processed distinctively, influenced not just by their emotional charge but also by their broader social and communicative connotations[7].

**Conclusion and Limitations**

The present study was aimed to exploit the potential of machine learning methods to decode word categories from EEG responses. Our results demonstrate the model's capacity to distinguish between neutral, taboo, and negative stimuli under Look and Accept conditions. In both Look and Accept conditions, the SVM model displayed above chance accuracy in classifying stimuli. The involvement of specific cortical regions and time windows, notably the central-parietal-occipital and anterior right scalp regions, reveals the complex neural mechanisms involved in emotional word processing. These findings align with established distribution and timing of the LPP responses to emotional stimuli. Interestingly, the Accept condition demonstrated slightly reduced accuracy but still significant. Moreover, our study provides novel insights into the distinct processing of taboo words. Despite their higher valence compared to negative words, taboo words elicited higher ERP responses and accuracy in classifications. This highlights the unique cognitive and neural processing of taboo words, possibly due to their social and contextual significance, which adds a layer of complexity beyond mere emotional valence.

Nonetheless, this study has some limitations that should be noted. First, the use of word stimuli could limit the broader applicability of our results. To enhance the validity and applicability of the results, it is recommended that future studies should include other stimuli, such as pictures. Furthermore, decoding algorithms are designed to identify distinguishing patterns between classes in the data. However, unlike traditional brain imaging analyses that aim to localize specific brain regions responsible for the signal, decoding methods do not always reveal the precise neural sources of these patterns.

**Methods**

**Participants**

Forty native Italian speakers (16 males; mean age = 23.78 years, SD = 3.40; mean years of education = 15.64, SD = 2.01) participated in the study. All participants were right-handed with normal or corrected-to-normal vision, no auditory impairments, and no reported neurological, psychiatric, or learning disabilities. The sample size was chosen based on prior EEG studies focused on emotion regulation using word stimuli[20] . Three participants were excluded due to the excessive number of noisy epochs (> 25%), and two were removed due to artifacts induced by a malfunctioning EEG cap, leaving 35 participants in the final sample. All participants signed an informed consent document and were compensated with €15 for about three hours of involvement, including preparation and experimental procedures. The study procedure received approval from the university of Trento research ethics committee protocol number (2021-033).

**Stimuli**

Eighty neutral and 80 negative words were selected from the Italian adaptation of the Affective Norms for English Words (ANEW[48,49]). Additionally, 80 taboo words were sourced from the Italian Taboo Words database (ITABOO[50]). Words in the three categories were comparable with respect to the psycholinguistic variables shown in Table 1. For purposes of counterbalancing, each category of stimuli was split into two subsets, each containing 120 items.

…………………………………………………..

Please insert Table 1 about here

…………………………………………………..

**Apparatus and Procedure**

*Training*

Prior to the EEG recording, a comprehensive training session was conducted by the experimenter. This session combined detailed written and verbal instructions to ensure that participants were able to implement the acceptance strategy during the experiment. Participants practiced this strategy with 10 unpleasant images from the International Affective Picture System (IAPS[51]). They were instructed to apply the acceptance strategy to each image. Additionally, participants evaluated their emotional responses in terms of valence and arousal dimension using the Self-Assessment Manikin (SAM) method [52], which employs independent 1 to 9-point Likert scales for each dimension.

*Task and Design Experiment*

Upon their arrival at the laboratory, participants filled out a consent form and provided demographic information, including age, gender, and years of education. They also confirmed having no auditory impairments, and no reported neurological, psychiatric, or learning disabilities. The EEG cap (64-channels) was then installed. After these preliminary steps, they received both verbal and written instructions before beginning the training. The experimental procedures and behavioral data collection were conducted using E-Prime 2 software (Version 2.0.10.356, Psychology Software Tools), which controlled the presentation of stimuli and the recording of responses. The study used a within-participants design consisting of two blocks, featuring the Look (i.e., no regulation) and the Accept (i.e., apply the acceptance regulation strategy) conditions. One set of words was administered in each block. Each set included 40 neutral, 40 negative, and 40 taboo words. The order of the two conditions (Look vs Accept) and the assignment of the 2 sets of words to each block was counterbalanced across participants. In each block, stimulus presentation was randomized for each individual.

Participants completed a total of 240 trials. Each trial began with a 1.5s fixation cross, followed by the word, displayed for 4s. Word stimuli were followed by a 0.5 s blank screen, after which participants first evaluated the valence and then arousal on a 1 to 9 scale using the standard SAM method. On this scale, a valence score of 1 indicated "very unpleasant," and 9 "very pleasant"; an arousal score of 1 meant "not arousing," and 9 "very arousing". Each trial finished with a 2s blank screen, serving as an inter-trial interval. Participants could take breaks after every 40 trials, at their discretion, to minimize fatigue and maintain comfort. (Fig.1). The entire experimental session, including the installation of the EEG cap, lasted approximately 150 minutes.

…………………………………………………..
Please insert Figure 1 here
…………………………………………………..

**EEG Data Acquisition**

EEG recordings were conducted using the eegosport (ANT) system at a sampling rate of 1000 Hz using 64 Ag–AgCl scalp electrodes arranged in the standard 10/10 layout. The CPz electrode was used as the online reference, and electrode impedances were kept below 20 kΩ to ensure adequate signal quality. Additionally, an electrooculogram (EOG) was obtained using electrodes positioned below the left eye. For signal processing, we used EEGLAB[53] (version 2021.1;) and FieldTrip toolbox[54] (version fieldtrip-20230913); together with custom scripts within the MATLAB environment (version 2021b, MathWorks Inc).

**EEG pre-processing**

The mastoid electrodes were not included in the analysis. The EEG data were first subjected to a 0.1 Hz high-pass filter, followed by an 80 Hz low-pass filter using a second-order Butterworth filter for both stages. The filtered continuous EEG signal was segmented into epochs extending from 1500 ms before to 4500 ms after the onset of stimuli. Noisy channels were interpolated using spherical interpolation ($M = 0.77$), and the data were re-referenced to the average activity across all electrodes. An Independent Component Analysis (ICA) was performed using the AMICA algorithm (Palmer et al., 2008), and components associated with eye blinks or saccades were removed ($M = 1.85 \pm 0.40$). To address potential rank-deficiency issues, the number of independent components was calculated while considering both the number of interpolated channels and the use of the average reference. Additional artifact rejection steps included the removal of noisy epochs where the signal exceeded ±100 µV in any channel, with an average of 6.66% epochs discarded. Participants were excluded from the EEG analyses when more than 25% of their data were removed. Average ERPs were computed for each condition and electrode.

Previous studies have extensively described the spatio-temporal dynamics of LPP modulations triggered by emotional words (e.g., [9]). However, the specific impact of these modulations within our experimental setup was less clear. We used two approaches to choose the scalp electrodes and time points for analysis. First, to identify the most relevant channels and time windows for in-depth analysis, we employed a cluster-based permutation analysis[55]. This analysis was aimed at comparing negative and neutral stimuli to pinpoint the precise temporal and spatial coordinates that would be most effective for later evaluations of emotion regulation strategies. We included all data from 0.4 to 1.5 s centered on stimulus onset, encompassing data from all electrodes. The analysis identified three significant clusters. Aggregation of these clusters helped in pinpointing key channels and timing for further study. Specifically, we focused on channels active for at least 50% of the cluster's duration and on samples in which at least 50% of the cluster's channels were involved. The channels selected for further analysis included P5, PO5, PO3, P3, P1,

CP3, Pz, POz, CP1, C3, CPz, C1, CP2, P2, and Cz. The critical time window for further analysis was determined to be between 0.637s and 0.878s (Region 10; see Table 2 and Figure 2).

Secondly, we adopted a more general approach, by segmenting the scalp into three main areas: anterior, central, and posterior. Each area was further divided into left, center, and right, resulting in nine distinct regions[56] (Regions 1-9, detailed in Table 2 and Figure 3). For each region, we computed the average activity across channels within two designated time windows of interest. These time windows were defined as the early (E; 450-649 ms) and the late (L; 650-850 ms) phases. We selected these specific intervals based on a visual inspection of ERP plots across different experimental conditions. Additionally, to investigate the effect of strategy independently from word type, we analyzed the 1500 ms pre-stimulus period by averaging the brain activity for each region across all stimuli. This allowed us to examine the Look and Accept conditions independent of word category.

Our decision to use these two approaches is based on their complementary nature. The data driven cluster-based permutation analysis provided a detailed roadmap, identifying specific regions and time windows where the maximum effects were found with respect to the classic contrast between negative and neutral words. It hence captures the experiment-specific spatiotemporal coordinates of the most frequently investigated contrast. This allowed us to target relevant areas highlighted by the extant literature for a more in-depth analysis. However, we also aimed to capture broader patterns of activity across the brain. The more comprehensive whole-brain approach ensured that we didn't overlook any smaller effect in the other regions of the brain that might extend beyond the specific clusters. Together, these methods provided both precision and a holistic view of brain activity, leading to a more thorough understanding of the data.

…………………………………………………..

Please insert Table 2 about here

…………………………………………………..

…………………………………………………..

Please insert Figure 2 about here

………………………………………………..…

…………………………………………………..

Please insert Figure 3 about here

…………………………………………………

**Behavioral analyses**

To investigate the differences between word categories, a series of paired-samples t-tests were conducted to compare valence and arousal ratings across neutral vs. taboo, neutral vs. negative, and negative vs. taboo in both the Look and Accept.

**Machine learning analysis**

We applied Support Vector Machine (SVM) classification model inside NeuroMiner toolbox (NM) version 1.05[57] in Matlab (v.2023). The objective of the model was to distinguish between neutral vs. negative, neutral vs. taboo, negative vs. taboo word stimuli separately within both the "Look" and "Accept" conditions and to investigate the effect of strategy independent of stimuli type. The model was developed using a repeated-nested double cross-validation (CV) method. This method is crucial for preventing information leakage, reducing the risk of overfitting, and providing an unbiased estimation of the model's generalizability to novel data[58–60]. This cross-validation structure included two nested k-fold CV cycles: an inner cycle (CV1) where models are built, and a superordinate outer cycle (CV2) where they are tested for generalizability[61]. Each cycle consisted of 10 folds and underwent 10 permutations. Preprocessing in CV1 involved removing zero-variance features and normalizing matrices to the mean[62].. Missing values were addressed using a k=7 nearest neighbor method for imputation[63]. The data processed in CV1 was then fed into a linear, class-weighted SVM algorithm (LIBSVM 3.1.2 L1-Loss SVM) to construct a hyperplane for making predictions in the training and test sets. The optimal analysis pipeline identified was subsequently applied to each k-fold and each permutation in the CV2 cycle. The classification of word stimuli (neutral, taboo; neutral, negative; negative, taboo) in both conditions was determined by a majority vote across all ensemble models. Permutation testing was conducted to establish statistical significance at an alpha level of 0.05 with 1,000 permutations[64]. The performance metrics for the model included Balanced Accuracy (BAC), sensitivity, specificity, Positive Predictive Value (PPV), Negative Predictive Value (NPV), and the Area Under the Curve (AUC). BAC provides an overall measure of the model's effectiveness, expressed as a percentage, balancing its ability to correctly classify both classes (e.g., neutral and taboo stimuli), especially when there is class imbalance. Sensitivity (or True Positive Rate) is expressed as a percentage and reflects the model's ability to correctly identify instances of one class, such as accurately classifying taboo stimuli when they are indeed taboo. Specificity (or True Negative Rate), also expressed as a percentage, indicates the model's effectiveness in correctly identifying instances of the other class, such as accurately classifying neutral stimuli when they are indeed neutral, thereby avoiding incorrect classifications. PPV is expressed as a percentage and indicates the proportion of instances classified as taboo that are indeed taboo, demonstrating how reliable the model is when it predicts a taboo outcome. NPV, similarly expressed as a percentage, represents the proportion of instances classified as neutral that are indeed neutral, showing the reliability of the model when predicting a neutral outcome. Lastly, the AUC provides a comprehensive measure of the model's overall ability to distinguish between the neutral and taboo classes across all possible thresholds; it is typically expressed as a value between 0 and 1, rather than a percentage, with higher values indicating better overall performance.

**Data Availability Statement**

The dataset generated and analyzed during the current study is available from the corresponding author upon reasonable request

**Author Contributions**

P.A.G.: Collected data, performed data analysis, contributed to the writing of the manuscript; M.S.: Provided supervision, reviewed the manuscript; B.M., P.W., R.J.: Reviewed the manuscript; A.G.: Conceived the work, performed data analysis, reviewed the manuscript, and provided supervision.

**Competing Interests**

The authors declare no competing interests.

Figure 1. Experiment Design

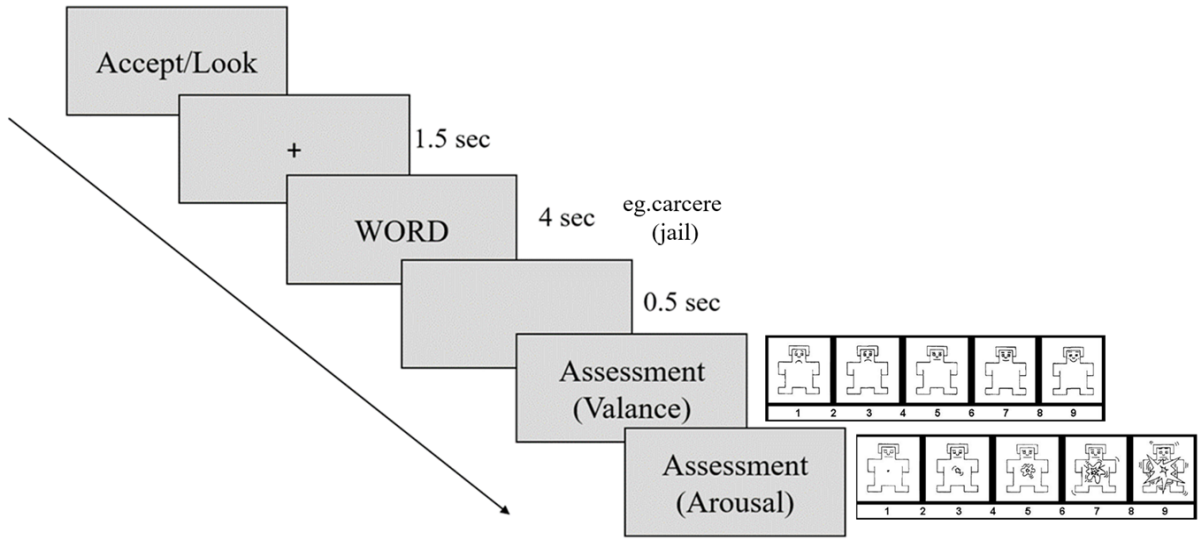

Figure 2: Electrode Grouping Layout Region 1-9

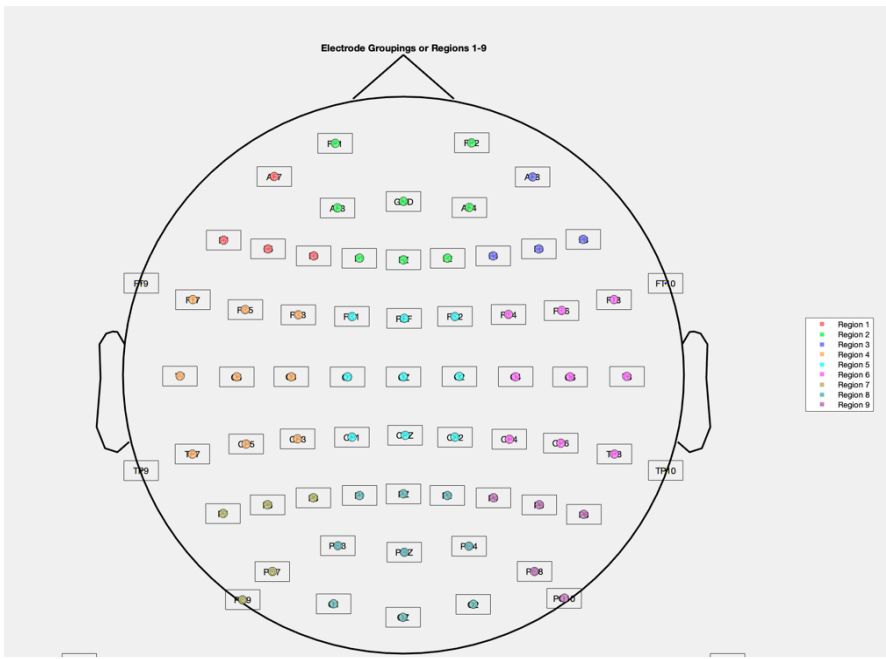

Figure 3: Electrode Grouping Layout Region 10

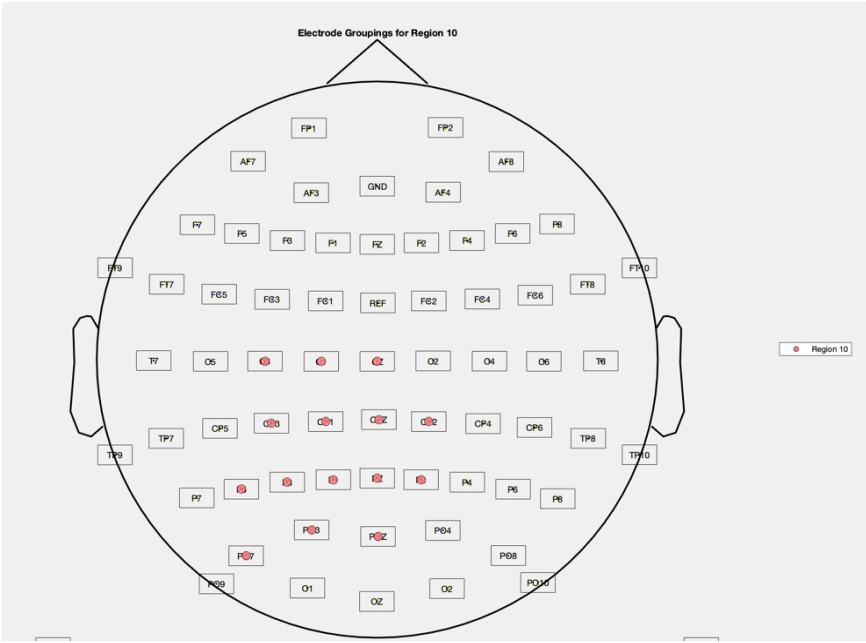

Figure 4. SVM classification performance in the Look condition. Top figure: SVM decision scores distinguishing Neutral (Neu) from 'Taboo (Ta) stimuli, with the confusion matrix and Receiver Operating Characteristic (ROC) curve displayed on the right. Middle figure: SVM decision scores for Neutral versus Negative stimuli, alongside the confusion matrix and ROC curve. Bottom figure: SVM decision scores for Negative and Taboo stimuli, with the corresponding confusion matrix and ROC curve.

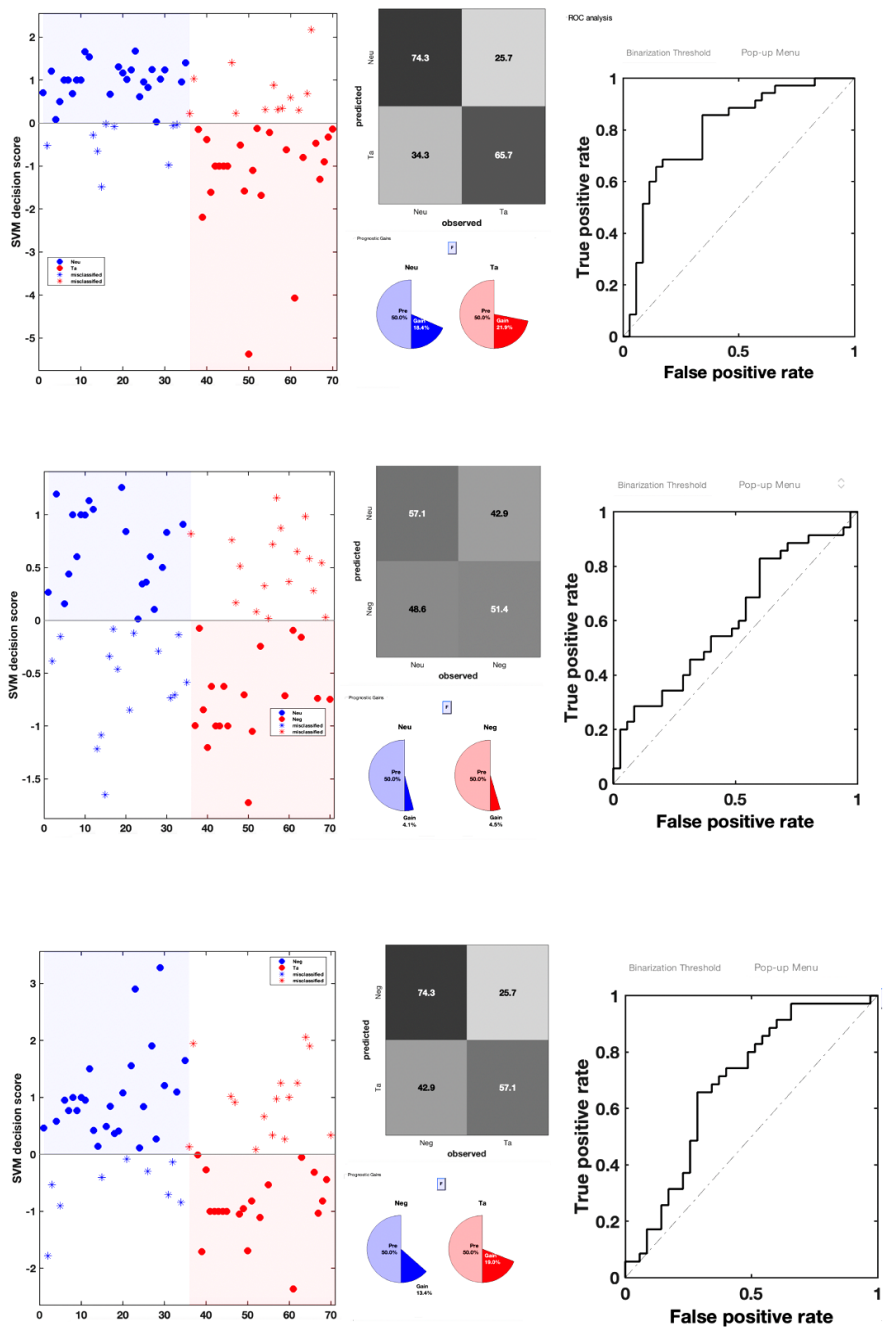

Figure 5. ERP plot of responses to word categories in the Look Condition. Top: Neutral vs. Taboo, Middle: Neutral vs. Negative, Bottom: Negative vs. Taboo.

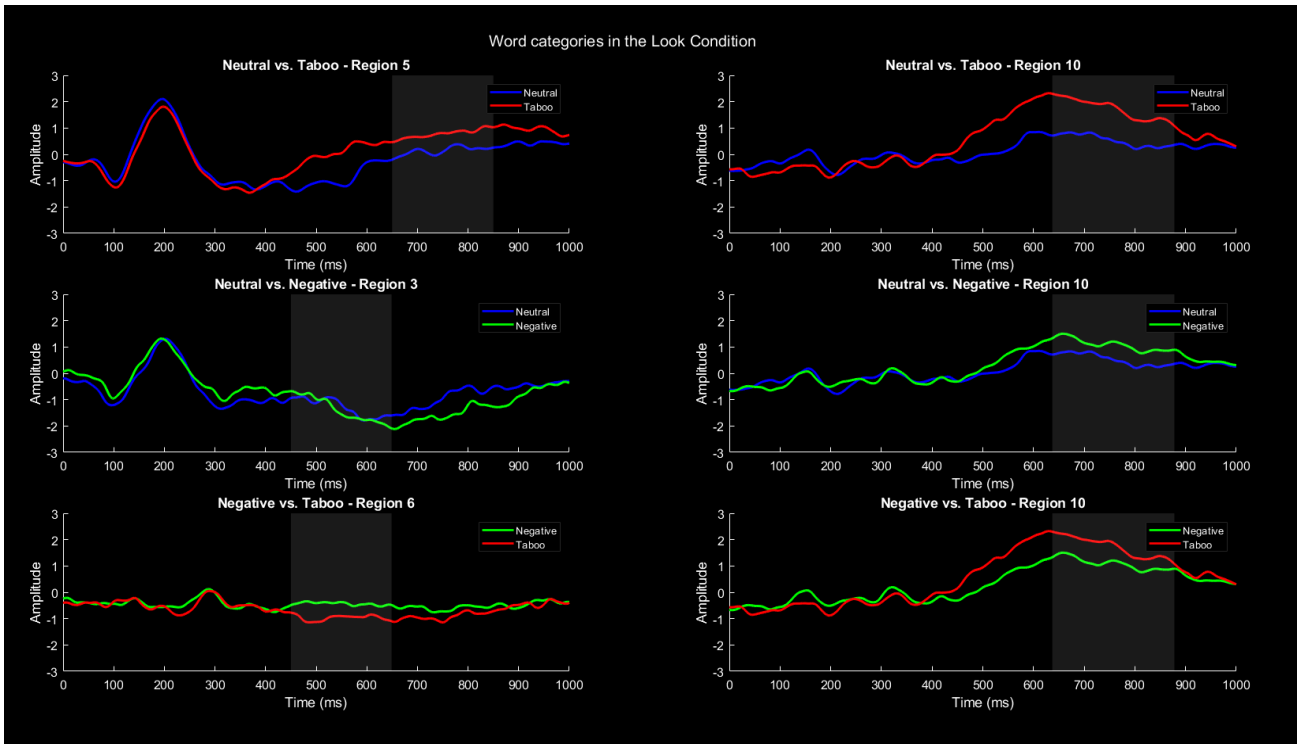

.

Figure 6. SVM classification performance in the Accept condition. Top panel: SVM decision scores differentiating 'Neutral (Neu) from 'Taboo (Ta) stimuli, with the confusion matrix and Receiver Operating Characteristic (ROC) curve on the right. Middle panel: SVM decision scores for 'Neutral (Neu)' versus Negative (Neg) stimuli, alongside the confusion matrix and ROC curve. Bottom panel: SVM decision scores for 'Negative (Neg) and Taboo (Ta) stimuli, with the corresponding confusion matrix and ROC curve

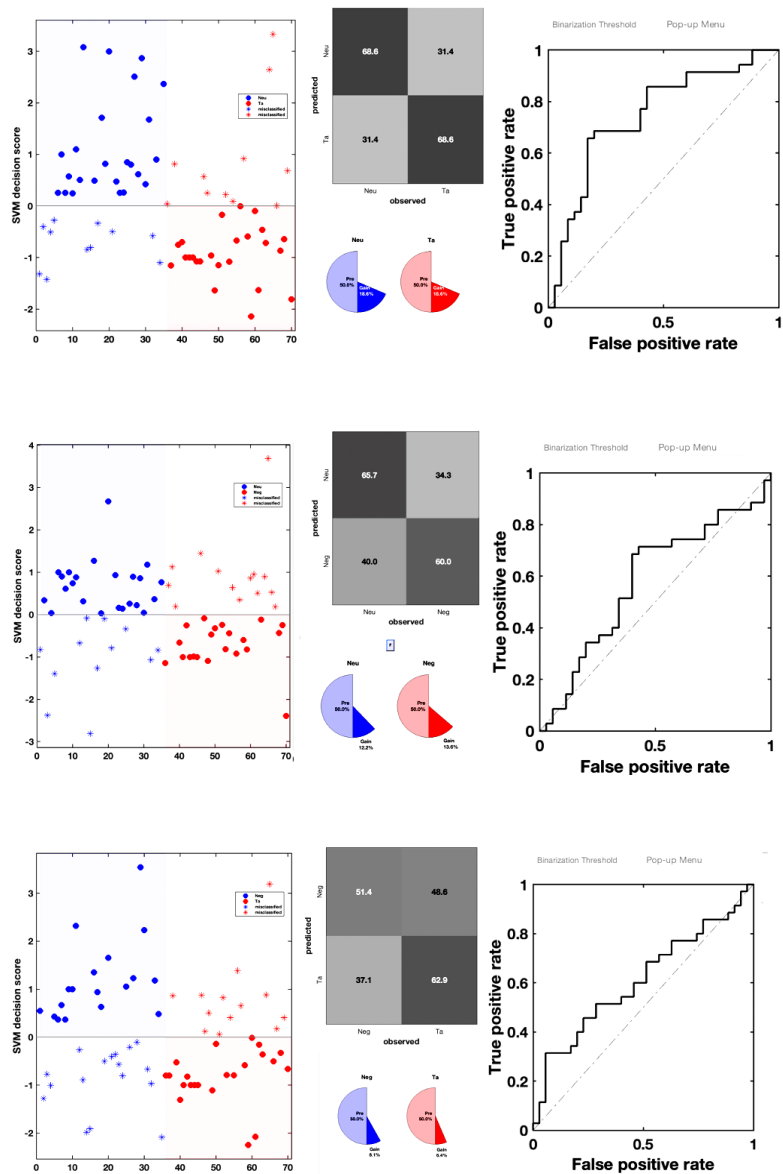

Figure 7. ERP plot of responses to word categories in the Regulate Condition. Top and second rows: Neutral vs. Taboo. Third row left: Neutral vs. Negative. Third row right and bottom row: Negative vs. Taboo

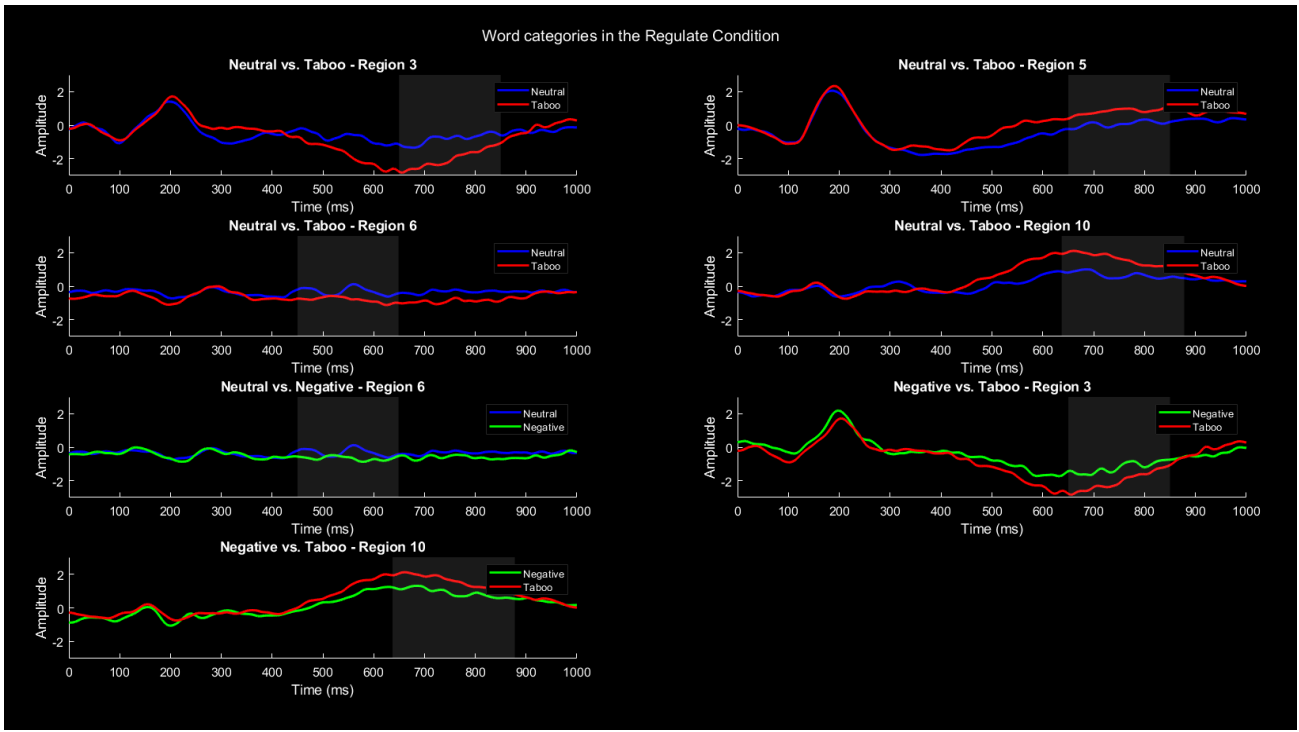

Figure 8. SVM classification performance in the Accept and Look condition. Top panel: SVM decision scores differentiating Accept(A) from Look (L) condition, with the confusion matrix and Receiver Operating Characteristic (ROC) curve on the right

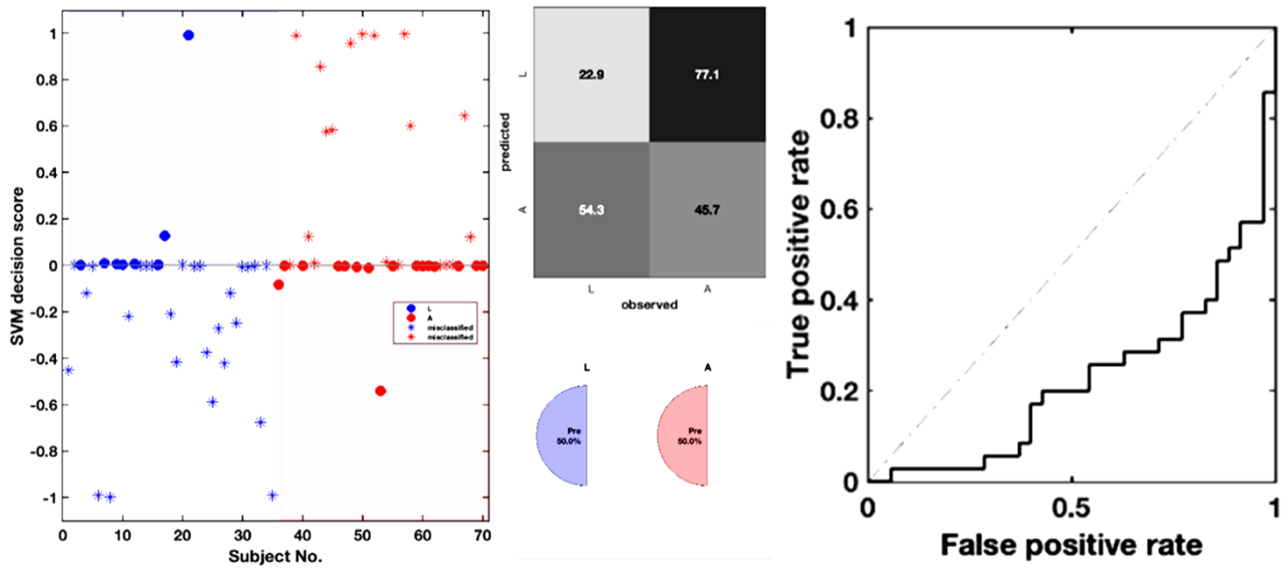

Table 1 Mean Psycholinguistic variables of the stimuli used in the experiment.

| Stimulus Type | Length of Letter | Orthographic Neighborhood | Familiarity | Concreteness | Valance | Arousal | Log of Frequency |
|---|---|---|---|---|---|---|---|
| Neutral | 7.30 ± 1.93 | 3.81 ± 5.32 | 5.84 ± 0.96 | 6.16 ± 1.70 | 5.69 ± 0.71 | 4.76 ± 0.68 | 6.04 ± 1.84 |
| Negative | 7.56 ± 1.87 | 3.29 ± 4.30 | 5.58 ± 1.03 | 6.03 ± 1.30 | 2.19 ± 0.38 | 6.32 ± 0.57 | 6.13 ± 1.76 |
| Taboo | 7.44 ± 1.79 | 4.81 ± 8.03 | 5.74 ± 1.16 | 6.46 ± 1.27 | 4.25 ± 1.03 | 5.01 ± 0.58 | 5.74 ± 1.89 |

Table 2; Electrode Groupings.

| Region | Laterality | Region | Electrodes |
|---|---|---|---|
| Anterior | Left | 1 | F3, F5, F7, AF7 |
| | center | 2 | FP1, FPZ, FP2, AFZ, AF4, AF3, F1, FZ, F2 |
| | Right | 3 | F4, F6, F8, AF8 |
| Central | Left | 4 | FC3, FC5, FT7, T7, C3, C5, T7, CP3, CP5, TP7 |
| | center | 5 | FC1, FCZ, FC2, C1, CZ, C2, CP1, CPZ, CP2 |
| | Right | 6 | FC4, FC6, FT6, C4, C6, T8, CP4, CP6, TP8 |
| Posterior | Left | 7 | P3, P5, P7, PO7, PO5 |
| | center | 8 | P1, PZ, P2, PO3, POZ, PO4, O1, OZ, O2 |
| | Right | 9 | P4, P6, P8, PO6, PO8 |
| | | 10 | P5, PO5, PO3, P3, P1, CP3, Pz, POz, CP1, C3, CPz, C1, CP2, P2, and Cz |

Table 3: Behavioural Result

| Condition | Variable | Comparison | $t$ | $df$ | $p$ | Mean difference | 95% CI Lower | 95% CI Upper |
|---|---|---|---|---|---|---|---|---|
| Look | Valence | Neutral vs. Taboo | 5.79 | 34 | < .001 | 0.92 | 0.6 | 1.24 |
| Look | Arousal | Neutral vs. Taboo | -8.65 | 34 | < .001 | -1.44 | -1.78 | -1.1 |
| Look | Valence | Neutral vs. Negative | 9.92 | 34 | < .001 | 1.93 | 1.54 | 2.33 |
| Look | Arousal | Neutral vs. Negative | -6.32 | 34 | < .001 | -1.61 | -2.12 | -1.09 |
| Look | Valence | Negative vs. Taboo | -13.19 | 34 | < .001 | -1.01 | -1.17 | -0.85 |
| Look | Arousal | Negative vs. Taboo | 1.11 | 34 | 0.28 | 0.16 | -0.13 | 0.46 |
| Regulate | Valence | Neutral vs. Taboo | 5.37 | 34 | < .001 | 0.9 | 0.56 | 1.24 |
| Regulate | Arousal | Neutral vs. Taboo | -9.06 | 34 | < .001 | -1.37 | -1.68 | -1.06 |
| Regulate | Valence | Neutral vs. Negative | 8.31 | 34 | < .001 | 1.88 | 1.42 | 2.34 |
| Regulate | Arousal | Neutral vs. Negative | -6.55 | 34 | < .001 | -1.52 | -2 | -1.05 |
| Regulate | Valence | Negative vs. Taboo | -7.80 | 34 | < .001 | -0.98 | -1.24 | -0.73 |
| Regulate | Arousal | Negative vs. Taboo | 1.12 | 34 | 0.27 | 0.15 | -0.13 | 0.43 |

*Note*: df: Degrees of freedom, CI: confidence interval

Table 4; Performance of SVM Model Across Different Conditions and Stimulus Types

| Condition | Stimulus Comparison | Balanced Accuracy | Sensitivity | Specificity | PPV | NPV | AUC | p value |
|---|---|---|---|---|---|---|---|---|
| Look | Neutral vs. Taboo | 70.00% | 74.30% | 65.70% | 68.40% | 71.90% | .80 | 0.01 |
| | Neutral vs. Negative | 54.30% | 57.10% | 51.40% | 54.10% | 54.50% | .60 | 0.0316 |
| | Negative vs. Taboo | 65.70% | 74.30% | 57.10% | 63.40% | 69.00% | .70 | 0.01 |
| Accept | Neutral vs. Taboo | 68.60% | 68.60% | 68.60% | 68.60% | 68.60% | .70 | 0.01 |
| | Neutral vs. Negative | 62.90% | 65.70% | 60.00% | 62.20% | 63.60% | .60 | 0.01 |
| | Negative vs. Taboo | 57.10% | 51.40% | 62.90% | 56.40% | 57.70% | .60 | 0.01 |
| Accept vs Look | Independent of Stimuli type | 34,30% | 24,30% | 45,70% | 36,20% | 37,20% | 0,22 | |

*Note*: PPV = Positive Predictive Value, NPV = Negative Predictive Value, AUC = Area Under Curve.